\documentclass[12pt]{iopart}
\eqnobysec
\usepackage{graphicx}

\font\amsb=msbm10
\def\hbar{\mbox{\amsb\char'175}}

\newcommand{\be}{\begin{equation}}
\newcommand{\ee}{\end{equation}}

\newcommand{\vecI}{{\mathbf I}}
\newcommand{\vxi}{{\mathbf \xi}}
\newcommand{\x}{{\mathbf x}}

\newcommand{\y}{{\mathbf y}}

\newcommand{\J}{{\mathbf J}}

\newcommand{\C}{{\mathbf C}}
\newcommand{\K}{{\mathbf K}}

\newcommand{\opA}{{\widehat{A}}}
\newcommand{\opB}{{\widehat{B}}}

\newcommand{\opR}{{\widehat{R}}}
\newcommand{\opT}{{\widehat{T}}}
\newcommand{\oprho}{{\widehat{\rho}}}

\newcommand{\opH}{\widehat{H}}
\newcommand{\opU}{\widehat{U}}
\newcommand{\opI}{\widehat{I}}

\newcommand{\M}{{\mathbf M}}

\newcommand{\zero}{{\mathbf 0}}

\newcommand{\der}{\partial}

\newcommand{\GO}{{\mathcal O}}

\begin{document}

\title{On the semiclassical evolution of quantum operators}

\author {A. M. Ozorio de Almeida\footnote{ozorio@cbpf.br} and O. Brodier}
\address{Centro Brasileiro de Pesquisas Fisicas, 
Rua Xavier Sigaud 150, 22290-180, 
Rio de Janeiro, R.J., Brazil.}

\begin{abstract}
The Heisenberg evolution of a given unitary operator corresponds classically
to a fixed canonical transformation that is viewed through a moving
coordinate system. The operators that form the bases of the Weyl representation 
and its Fourier transform, the chord representation, are, 
respectively, unitary reflection and translation operators. 
Thus, the general semiclassical study of unitary operators 
allows us to propagate arbitrary operators, including density
operators, i.e. the Wigner function. The various propagation kernels
are different representations of the superoperators 
which act on the space of operators of a closed quantum system. 
We here present the mixed semiclassical
propagator, that takes translation chords to reflection centres, or vice versa.
In contrast to the centre-centre propagator that directly evolves Wigner functions,
it is guaranteed to be caustic free, having a simple WKB-like universal form for a finite time, 
whatever the number of degrees of freedom.
Special attention is given to the {\it near-classical region} of small chords, since this dominates
the averages of observables evaluated through the Wigner function.

\end{abstract}

\maketitle

\section{Introduction} 

The semiclassical theory for the evolution of the states of a closed quantum system has a long history.
The way in which operators propagate semiclassically driven by the Heisenberg equation
is a more delicate matter. On the one hand, observables that give rise to smooth,
approximately classical functions in the Weyl representation 
(admissible operators \cite{Vor76}) also propagate in an almost
classical manner. In contrast, unitary operators exhibit semiclassically narrow oscillations
in the Weyl representation and so do density operators, represented by the Wigner function.
The propagation of these oscillatory regions is not trivially related to the corresponding
classical evolution, but it has been shown that the semiclassical Wigner function is
transported by pairs of phase space trajectories \cite{RiosOA}. Alternatively, 
these may be combined into a single trajectory in {\it double phase space}, 
secondary phase space \cite{OsKon}, or, in full rigor, the symplectic groupoid \cite{Rios04}.  

A general qualitative picture in terms of a pair of orbits can always be invoked.
If the {\it chord} that connects the initial points of the relevant 
pair of orbits is small, we may substitute these by a single central orbit, 
so that the propagator for Wigner functions assumes 
the limiting $\delta$-function form along the classical trajectory, 
deduced semiclassically by Marinov \cite{Mar91}. 
But, in the case of long chords, 
both the orbits at the chord tips must be used. 
The dificulty is that the initial chord depends on
the initial state, so that it is not so easy to derive a general 
semiclassical propagator. Furthermore, the limit of small chords is a caustic
of the semiclassical theory demanding a higher uniform approximation \cite{Dittrich} than 
that developed in \cite{RiosOA, OsKon}.

The Weyl, or centre representation decomposes arbitrary operators into a superposition
of reflection operators \cite{Grossmann}. The Fourier transform of the Weyl symbol, i. e.
the chord symbol, is the expansion coefficient of the same operator in 
the basis of translation operators. Both these bases belong to the general class 
of unitary operators, so that the semiclassical theory for evolution
of unitary operators allows us to propagate arbitrary Weyl symbols, or Wigner functions.
The purpose of this paper is to provide this general semiclassical framework.
If one considers that the space of linear operators that act on Hilbert space 
also forms a linear space, then the kernels of the integral representation
for the propagation of Weyl symbols and chord symbols correspond to representations
of {\it super-operators} \cite{Breuer}. Here, we are only concerned with a subclass of super-operators,
or non-selective operations \cite{Kraus}, appropriate to closed quantum systems, but a generalization
of the present semiclassical theory to open Markovian systems will soon follow. 

The natural setting for the present semiclassical theory is double phase space, 
whose elements are all the ordered pairs of phase space points and hence encompasses
all possible classical transitions. A uniform translation of the ordinary phase space
is represented by a plane in double phase space, 
which is transverse to the double phase space plane 
that defines a canonical reflection through a phase space point. 
The points on these alternative sets of planes
can be used as conjugate coordinates for double phase space, which correspond,
respectively, to the chord and the centre (Weyl) representations. It is the classical
evolution of these planes in double phase space that supplies the semiclassical form
of the propagators that represent super-operators.

The following section reviews the semiclassical correspondence between 
classical canonical transformations and quantum unitary operators. 
The choice of a particular representation for the latter is necessarily paired
to a specific type of generating function for the corresponding classical transformation. In the case
of the centre, or the chord representations, the respective centre and chord
generating functions are subject to caustics wherever the canonical transformation 
can be locally approximated by either a reflection, in the case of 
the centre generating function, or a translation \cite{Alm98}. These give rise
to spurious singularities in the simplest semiclassical approximations
that are reviewed in section 3, which also presents the corresponding
forms of the Hamilton-Jacobi equation appropriate to both these generating functions.

Up to this point, the evolution of the unitary operators results from solving the Schroedinger
equation. It is only in section 4 that the less studied case of the Heisenberg
evolution is reviewed with reference to the Weyl propagator and its corresponding
Hamilton-Jacobi equation. No distinction will be made here with the von Neumann
equation, appropriate to density operators that are evolved unitarily, 
because it is simply related to the Heisenberg equation by a time inversion, 
or by changing the sign of the Hamiltonian.
A direct semiclassical 
correspondence follows for this type of quantum evolution in the case of unitary
operators, if we view a fixed classical canonical transformation 
from a moving (canonical) coordinate frame.

The notion of double phase space is introduced in section 5, within the general
point of view developed by Amiet and Huguenin \cite{AmHu80, Rios04}. As well as
providing an appealing reinterpretation of the previous results, this allows
the simplest derivation of the Hamilton-Jacobi equation for the chord
generating function. A revised study of the general evolution of operators
follows in section 6.

Section 7 is dedicated to the semiclassical evolution of translation operators.
They correspond to planes in double phase space that are analogous to the momentum planes in
ordinary phase space. Hence, the chord symbol is a $\delta$-function, whereas the Weyl
(centre) symbol is a plane wave. The latter is the preferred basis in which to
view the general nonlinear semiclassical evolution of tranaslations into a single WKB-like wave. 
The Weyl symbols of translation operators are propagation kernels
for the evolution of arbitrary operators that are specified initially by their
chord symbol and finally by their Weyl symbol. The invariance of the identity
operator under Heisenberg evolution corresponds to the classical invariance of the
zero-chord plane. It is only within this plane that the classical 
double phase space motion generally coincides with the classical Liouville flow in 
ordinary phase space.

In section 8 we examine the evolution of the conjugate reflection operators.
These correspond to planes in double phase space which are analogous to 
the position planes in ordinary phase space. Hence, the Weyl symbol is a 
$\delta$-function, whereas here it is the chord symbol that evolves 
into a single nonlinear WKB-like wave from an initial plane wave. 
Therefore, it is guaranteed to have 
no interference and no caustics for a finite time. 
The chord symbol for the reflection operator is the
propagation kernel for initial Weyl symbols of arbitrary operators 
to be later specified by their chord symbols. Of course, 
the final Weyl symbol, or Wigner function, which is the subject of
parallel work \cite{Dittrich}, can be obtained at any time 
by a Fourier transform. An example is provided of evolution driven by
a homogeneous cubic Hamiltonian. This is the simplest nontrivial case,
since the evolution for quadratic Hamiltonians reduces to the linear Liouville flow.

Our conclusions in the final section emphasise that actually we are dealing  with
a single mixed propagator, in spite of the alternative definitions that correspond to
the motion of alternative planes in double phase space. Its simple semiclassical form
is in sharp contrast to the direct centre-centre propagator for Wigner functions, 
whose caustic structure is analyzed in the Appendix.

\section{Correspondence between canonical and unitary transformations}

In this section we recall several known facts about classical-quantum correspondence
of unitary operators and canonical transformations. This generally only holds within a semiclassical
approximation, but the reflection and translation operators, which form respectively the bases of
the Weyl representation and its Fourier transform, belong to the special class for which
the correspondence is exact. 

Time dependent unitary operators that act on quantum states in Hilbert space correspond classically
to evolving canonical phase space transformations. In the case of motion generated by a 
constant Hamiltonian operator, $\opH$, the continuous group of unitary operators, 
\be
\opU_t= \exp{(-it\opH /\hbar)},
\label{opevol}
\ee
is related to the continuous group of canonical transformations $\C_t$. Indeed, if we 
define the points in the $2L$-dimensional phase space as $x=(p_1,...,p_L,q_1,...,q_L)$,
we have $\C_t:x_0\rightarrow x_t$, driven by the classical Hamiltonian $H(x)$
according to Hamilton's equations,
while the Hilbert space vectors evolve linearly:
$|\psi_t\rangle=\opU_t|\psi_0\rangle$. 
The various representations of the unitary operators correspond
to different generating functions. Semiclassically, the latter determine the phase of the 
corresponding quantum propagators. The Schroedinger equation for $\opU_t$,
e.g. in either of the position, momentum, or the Weyl-Wigner representations, corresponds to 
alternative versions of the Hamilton-Jacobi equation (see e.g.\cite{livro}). 

Even an initially single valued generating function may become multiplevalued as it evolves.
Its different branches are then connected along caustics, which lead to spurious singularities 
of the semiclassical approximation for the various representations of $\opU_t$. 
For this reason, prescriptions for transforming between the different representations 
is an important part of the semiclassical theory \cite{Maslov}. 
In the case of the Weyl propagator, $U_t(\x)$,
that is the Weyl-Wigner representation of $\opU_t$, the initial generating function
corresponding to the identity operator $\widehat{I}$, is just
$S_t(\x)=0$, so that the appearence of caustics is optimally delayed. Even so, 
they may eventually appear \cite{Alm98}, occuring where $\C_t$ can be locally approximated 
as a reflection about a point $\x$, i.e. the transformation
\be
R_{\x}: x_{-} \rightarrow x_{+}=-x_- +2\x.
\label{refl}
\ee

It is then appropriate to switch to the complementary phase space representation,
which, for an arbitrary operator, $\opA$, takes the form
\be
A(\vxi) = \frac{1}{\left(2\pi\hbar\right)^L} \int \rmd\x ~ 
\exp{\left(-\frac{i}{\hbar}\vxi\wedge\x\right)} A(\x).
\label{chord-prop}
\ee
Here we have used the skew product,
\be
x\wedge x'=\sum_{n=1}^L (p_l q'_l - q_l p'_l)= \J\>x \cdot x', 
\label{squew}
\ee
which also defines the skew symplectic matrix $\J$.
In the case of unitary operators, the Fourier transform between $U_t(\x)$
and $U_t(\vxi)$ corresponds classically to a Legendre transform between
$S_t(\x)$ and the new generating function, $S_t(\vxi)$.  
This describes the evolution in terms of uniform translations , 
\be
T_{\vxi}: x_{-} \rightarrow x_{+}= x_- +\vxi,
\ee
by a vector $\vxi$, which is a chord in the classical trajectory,
hence this is termed the chord generating function. 

Corresponding to this, 
the chord representation \cite{Alm98}, $A(\vxi)$, of an operator, $\opA$, 
can be defined directly as 
a superposition of the unitary translation operators
\be
\opT_{\vxi}= \exp {\left({i\over \hbar} \vxi \wedge \widehat x\right)},
\label{translation}
\ee
where $\widehat x=(\widehat p,\widehat q)$, that is,
\be
\opA= \int \frac{\rmd\vxi}{(2\pi\hbar)^L} \>A(\vxi) \> \opT_{\vxi}.
\label{chordrep}
\ee
Recalling the definition of the unitary reflection operators as
\be
2^L\opR_{\x}=\int \frac{\rmd\vxi}{(2\pi\hbar)^L} \>\opT_{\vxi} \exp({i\over\hbar}\x\wedge\vxi),
\label{quantreflection}
\ee
and comparing with (\ref{chord-prop}), we see that the Weyl representation, $A(\x)$, decomposes
an arbitrary operator, $\widehat A$, into a superposition of reflection operators \cite{Grossmann}:
\be
\opA= \int \rmd\x \>A(\x) \> 2^L\opR_{\x},
\label{Weylrep}
\ee
Probably the first to remark on the general structure of translations and reflections
underlying the Weyl and the chord representations were Grossmann and Huguenin \cite{GrossHug}.

The greater familiarity of the Weyl-Wigner representation
is justified by its use for the density operator, $\widehat{\rho}$, which
is represented by the celebrated Wigner function \cite{Wigner}, $W(\x)=\rho(\x)/(2\pi\hbar)^L$,
a real (but not necessarily positive) quantum quasiprobability in phase space. 
In the case of the density operator, its Fourier 
transform can be interpreted as a kind of quantum characteristic 
function, $\chi(\vxi)$, called in \cite{AlmVal04} simply the chord function.
The evolution of density operators in open systems is constrained by trace
conservation and positivity \cite {Kraus, Giulini}, 
but is not limited to the simple form that is studied
in this paper. It should be noted, however, that the basis 
operators for the Wigner function, namely the reflection operators, $\opR_{\x}$,
are not themselves positive, having degenerate eigenvalues $\pm 1$.

\section{The semiclassical propagators}

\begin{figure}
\includegraphics[width=8cm]{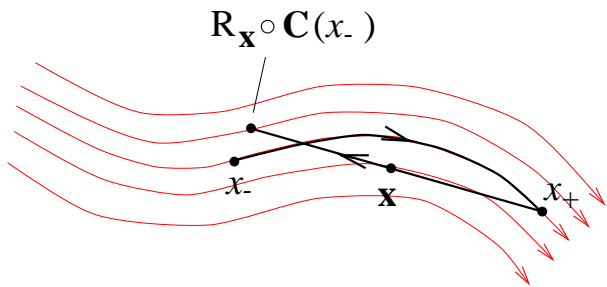}
\caption{The point $x_-(\x)$ is mapped onto $x_+(\x)$ by the reflection $R_\x$ in the same
way as it is moved by the given  canonical transformation $\C$. 
It follows that $x_-(\x)$ is a fixed point of the transformation $R_\x \circ \C$,
in which case the circuit $\gamma_t$ is defined by the composition of both these motions.}
\label{fig1}
\end{figure}

A given centre generating function, $S_\C(\x)$, defines the canonical transformation, 
$\C:x_-\rightarrow x_+$, implicitly through the equation,
\be
\frac{\der S_\C}{\der \x}= \J \>\vxi_\C(\x),
\label{dSdX}
\ee
with $\vxi_\C(\x)=x_+-x_-$, i. e. the chord centred on $\x$ \cite{AmHu80, Alm98}. 
This is subject to the usual constraints for generating functions \cite{Goldstein}.
It was shown by Marinov \cite{Mar79} that this centre action is just
\be
S_t(\x)= A_t(\x)-t H(x_-),
\label{centaction}
\ee
where $A_t(\x)$ is the symplectic area 
\be
A_t(\x)=\oint_{\gamma_t}p\cdot \rmd q
\ee
for the circuit $\gamma_t$ that starts along a trajectory segment 
chosen so that its endpoints are centred
on $\x$ and then is closed by reversing $\vxi_t(\x)$, as shown in Fig.\ref{fig1}.
Of course, for a Hamiltonian system, we have $H(x_+)=H(x_-)$, but generally $H(\x)\neq H(x_-)$.
It might appear awkward to find the point $x_-(\x)$, but this is merely the fixed 
point of the canonical transformation $R_\x \circ \C$, which
shows this to be a {\it reflection representation} 
of classical mechanics \cite{Alm98}, as shown in Fig.\ref{fig1}.
For short times, the trajectory segment is approximately $\vxi_t(\x)$ itself,
so that, like the trajectory, the chord is unique. In this limit, the first term in (\ref{centaction})
is negligible, resulting in the explicit approximation
\be
S_t(\x)\mathop{\rightarrow}_{t\rightarrow\zero} -t H(\x).
\label{smallt}
\ee

Linearizing the canonical transformation in the neighbourhood of the tips
of the chord centred on $\x$, so as to define the tangent map,
\be
\delta x_+ = \M(\x) \>\delta x_-,
\label{linearize}
\ee 
we obtain the Hessian matrix of $S(\x)$ as
\be
\frac{\der^2 S}{\der \x \der \x}= - \J [1-\M][1+\M]^{-1},
\label{Cayley}
\ee
which is a {\it Cayley parametrization} of $\M$ \cite{AmHu80, Alm98}.
References \cite{Alm98, Rios04} discuss some of the  properties and a few examples
of the centre generating function. It follows from (\ref{Cayley}) that the
{\it centre caustics} for the generating function are the manifolds where $\M(\x)$
has an eigenvalue $-1$, so that locally $\C$ is a reflection 
for one of the degrees of freedom.

Let us now consider the evolution equation of a centre generating function,
$S_t(\x)$, driven by a constant Hamiltonian, $H(x)$. 
Fixing the initial phase space point $x_-$ from which evolves 
$x_+(t)=\C_t(x_-)$, implies that $\delta S_t=-H(x_-)\delta t$, because of the centre
variational principle \cite{Alm90, Alm98}. 
Recalling that $x_- = \x-\vxi/2$ together with (\ref{dSdX}),
leads then to Marinov's version of the Hamilton-Jacobi equation \cite{Mar79}:
\begin{equation}
\frac{\partial S_t}{\partial t} + H \left(\x+\frac{1}{2}\J \frac{\partial
S_t}{\partial \x}\right) = 0.
\label{HJX}
\end{equation}
Several other partial differential equations will be presented in this paper
that describe the evolution of generating functions, in different contexts 
and depending on different free variables. We here follow the tradition of
naming all of these anachronically as Hamilton-Jacobi equations. 

These are the basic ingredients for constructing the semiclassical Weyl propagator,
\be
U(\x) = 2^L |\det (1+{\M})|^{-\frac{1}{2}} \exp [i\hbar^{-1}S(\x)+i\theta].
\label{SCUX}
\ee
Here, it makes no difference if the unitary operator, $\opU$, with Weyl symbol, $U(\x)$,
corresponds to a given canonical transformation $\C$, or a time dependent transformation
generated by a Hamiltonian $H(x)$.
The original derivation \cite{Berry89a} of $U(\x)$ in the latter case made no reference to
Marinov's generating function, but (\ref{SCUX}) follows by inserting (\ref{centaction})
in Berry's result. An alternative derivation based on the  Weyl path integral
is presented in \cite{Alm98}.  As usual, the semiclassical propagator 
is exact in the case of quadratic Hamiltonians, i. e. linear classical motion.

For a unitary transformation, $\opU_t$, that evolves continuously from the identity,
$\opI$, driven by $\opH$, we obtain from (\ref{smallt}) and the fact that $U_0(\x)=1$
that the phase $\theta =0$ for small times. 
This is only altered when a centre caustic of the generating
function $S_t(\x)$ is reached, so that the denominator of (\ref{SCUX}) blows up. 
But in this event, the eigenvalues of $\M$ that become $-1$ must come in pairs,
because this is a symplectic matrix (see e. g. \cite{Alm98}). Thus the change of
phase can only be a multiple of $\pi/2$, instead of just $\pi/4$, which is more usual for caustic
traversals. Beyond the passage of the first caustic through a given point $\x$, 
the centre generating function is no longer univalued,
so that the semiclassical Weyl propagator becomes a superposition of terms like (\ref{SCUX}),
one for each branch of the generating function. Of course, close to the caustic, the spurious
singularity of the semiclassical amplitude points to the need for 
an improved uniform approximation. One way to derive this is to transform back from the chord
propagator that will now be examined. 
This is the natural procedure from the double phase space point of view that
will be discussed in section 5. 

We can now obtain the semiclassical chord propagator by inserting (\ref{SCUX})
into (\ref{chord-prop}), that is
\be
U(\vxi) = \frac{1}{\left(\pi\hbar\right)^L} \int \rmd\x ~  
\frac {\exp [i\hbar^{-1}(S(\x)-\vxi\wedge\x)+i\theta]}{|\det (1+{\M}(\x))|^{\frac{1}{2}}},
\label{SCUxi1}
\ee
and then evaluating this Fourier integral by the method
of stationary phase. The stationary phase condition simply picks out the centre 
$\x(\vxi)$, such that $\vxi(\x)$ given by (\ref{dSdX}) equals the given chord $\vxi$.
The amplitude factor for stationary phase integration is then obtained from the
determinant of (\ref{Cayley}) as
\be
U(\vxi) = |\det (1-{\M})|^{-\frac{1}{2}} \exp [i\hbar^{-1}S(\vxi)+i\phi],
\label{SCUxi2}
\ee
where we identify the {\it chord generating function}, $S(\vxi)$, as the Legendre transform
of $S(\x)$. Indeed, it describes the same canonical transformation $\C$ by the relation \cite{Alm98},
\be
\frac{\der S_\C}{\der \vxi}= -\J \>\x_\C(\vxi),
\label{dSdxi}
\ee 
which is conjugate to (\ref{dSdX}). As discussed in \cite{Alm98},
we determine $x_-(\vxi)=\x(\vxi)-\vxi/2$ as the fixed point of $T_{-\vxi}\circ \C$
by a construction that is similar to Fig.\ref{fig1}.
Actually, there is a conjugate Cayley relation
between the matrix $\M$ that linearizes the canonical transformation and the Hessian matrix
of the chord action:
\be
\frac{\der^2 S}{\der \vxi \der \vxi}= - \J [1+\M][1-\M]^{-1}.
\label{Cayley'}
\ee
Thus we see that the spurious semiclassical singularity of $U(\vxi)$ in (\ref{SCUxi2}) takes place
at the caustic of the chord generating function where $\det[1-\M]=0$. This occurs when 
the canonical transformation in the neighbourhood of the chord tips can be
approximated by a uniform translation, or the identity.
Thus, the chord generating function is singular
as $t\rightarrow 0$ for a Hamiltonian flow. In contrast, the chord propagator
is perfectly regular at reflections, which correspond to the caustics of the
Weyl propagator. Hence, the complementarity of this pair of phase space representations.

It has been seen that the phase $\theta$ in (\ref{SCUX}) is zero for short times.
Therefore, the phase $\phi$ in (\ref{SCUxi2}) will be initially $-\pi/4$ times the
signature of the Hessian matrix for $H(\x(\vxi))$ see e.g. \cite {livro}). In the case of the simple 
quadratic Hamiltonians, this signature equals 2, for a harmonic oscilator $(L=1)$
and it equals 0, for the inverted oscillator. The explicit form of $S_t(\vxi)$ for these cases
is discussed in \cite{Alm98}. For quadratic Hamiltonians, the semiclassical chord propagator
is exact just as the Weyl propagator.

\section{Operator evolution}

There is another kind of evolution which may be imposed on a unitary operator,
taking along with the Weyl and the chord symbols.
This is just the usual evolution according to Heisenberg's equation,
\be
\frac{\der\widehat U'(t)}{\der t}=\frac{i}{\hbar}
\bigl[\widehat{H'}, \widehat U'(t)\bigr].
\label{Heisenberg}
\ee
Here, it is proper to distinguish the {\it external Hamiltonian}, $\widehat H'$,
from the Hamiltonian generator, $\widehat H$, which may have been used 
to define $\opU'(0)=\opU$ itself in (\ref{opevol}). 
Actually, (\ref{Heisenberg}) is more commonly employed for the propagation of observables
(such as, $\widehat q$, or $\widehat H$) or density operators (with a change of sign), 
rather than unitary operators.  Perhaps the more intuitive realization is to consider
$\opU$ as an active transformation, whereas $\widehat H'$ generates by exponentiation,
according to (\ref{opevol}), a continuous group 
of passive coordinate transformations in Hilbert space, $\widehat{V'}_t$.
Then we obtain 
\be
\opU'(t)=\widehat V'_{-t}\> \opU\>\widehat V'_{t}
\label{quantum}
\ee
as the time dependent unitary transformation that results from 
the adoption of the moving coordinate frame.
It is convenient to follow the nomenclature of Osborn and Kondratieva \cite{OsKon},
so as to distinguish the {\it Heisenberg evolution} of $\opU'(t)$ governed by (\ref{Heisenberg}) 
from the {\it Schroedinger evolution} of $\opU_t$. Though unusual, 
the latter is apropriate in as much as (\ref{opevol}) 
is the solution of the Schroedinger equation for the initial condition $\opU_0=\opI$. 

\begin{figure}
\includegraphics[width=8cm]{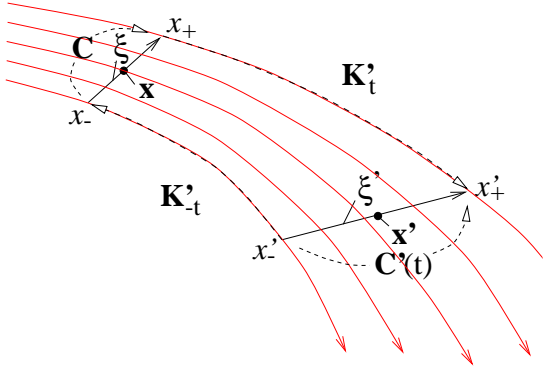}
\caption{The classical Heisenberg evolution, $\C'(t):\x'_-\rightarrow \x'_+$, 
of the initial canonical transformation, $\C:\x_-\rightarrow \x_+$, is constructed
with the addition of the forward orbit, $\K'_t:x_+\rightarrow x'_+$, and the backward
orbit, $\K'_{-t}:x'_-\rightarrow x_-$.}
\label{fig2}
\end{figure}

This specialization of the Heisenberg evolution to unitary operators
is the key to a clear classical correspondence. Indeed, we can also
view the classical canonical transformation, $\C:x_-\rightarrow x_+$, in the moving
coordinate frame defined by the continuous group of canonical transformations,
$\K'_t:x_0\rightarrow x_t$. The latter is generated 
by the classical Hamiltonian $H'(x)$, that corresponds to $\widehat H'$, not to the
generator of the transformation $\C$. Then the evolving
canonical transformation corresponding to $\opU'(t)$ is just
\be
\C'(t):x'_-\rightarrow x'_{+}=\K'_{-t}\circ \C \circ \K'_{t}(x'_{-}).
\label{classical}
\ee
Clearly, the evolution of the initial active transformation $\C$
depends here on a pair of orbits generated by the external Hamiltonian $H'(x)$, the
one for $x_+$ moving forwards in time, while the orbit for $x_-$ moves backwards
(or forwards for $-H'(x)$). This geometry is sketched in Fig.\ref{fig2}.
The fundamental difference between the {\it classical Heisenberg evolution},
$\C'(t)$, and $\C_t$ is highlighted by the example
where the initial active canonical transformation is just the identity: $\C=\vecI$.
No change of coordinate system can alter this, so $\C'(t)= \vecI$ for all time, 
which is not generally the case for the canonical transformation, $\C_t$, that results 
from the integration of Hamilton's equations.
Of course, the quantum operator $\opI$  is also invariant according to (\ref{quantum}),
in contrast to the nontrivial evolution specified by (\ref{opevol}).

The semiclassical Heisenberg evolution of operators in the Weyl representation was derived
independently in \cite{RiosOA} and in \cite{OsKon}.  
In both cases, the density
operator $\oprho(t)$ was considered, i. e. the propagation of the Wigner function, 
but the results can be immediately appropriated for the Weyl symbol of $\opU'(t)$,
with a change of sign for the time (strictly the von Neumann equation).
$U'(\x,t)$ has the same general form as (\ref{SCUX}), but with the alternative generating
function $S'(\x,t)$, corresponding to the canonical transformation $\C'(t)$ specified 
by (\ref{classical}) and the stability matrix $\M'(\x,t)$ obtained from \eref{Cayley}.  
The partial differential equation satisfied by $S'(\x,t)$ is then 
a kind of Hamilton-Jacobi equation \cite{OsKon},
\begin{equation}
\frac{\partial S'}{\partial t}(\x,t) 
+ H' \left(\x-\frac{1}{2}\J \frac{\partial S'}{\partial \x}\right) 
- H' \left(\x+\frac{1}{2}\J \frac{\partial S'}{\partial \x}\right) = 0.
\label{HJX2}
\end{equation}

It will be shown in the following section that the chord generating function
for the Heisenberg transportation of a canonical transformation satisfies a similar 
Hamilton-Jacobi equation. In the conclusion of the present section 
this same equation is anticipated through the study of the semiclassical limit 
of the Heisenberg equation in the chord representation. 
So as to represent the commutator in (\ref{Heisenberg}), we note that (see e.g. \cite{Alm98})
\be
\left(\opA\opB\right)(\vxi) = \frac{1}{(2\pi\hbar)^L}
\int d\vxi'~A(\vxi')B(\vxi-\vxi')
\exp{\left(\frac{i}{2\hbar}\vxi\wedge\vxi'\right)}.
\label{prod-rule-chord}
\ee
In the case of the Hamiltonian it is better to use the Weyl representation,
which is a smooth real function. On the other hand, linear phases can be
incorporated as a shift of the origin \cite{AlmVal04}:
\be
A(\vxi) \exp{\left(\frac{i}{2\hbar}\eta\wedge\vxi\right)}= 
A(\vxi-\eta).
\label{shift}
\ee
Hence, the chord commutator can be expressed as
\be
\fl[\opH',\opU'(t)](\vxi) = \frac{1}{(2\pi\hbar)^L}
\int d\vxi'd\x'~
\Biggl[  H'(\x'-\frac{\vxi}{2}) - H'(\x'+\frac{\vxi}{2})\Biggr]
U'(\vxi',t)
\exp{\left(-\frac{i}{\hbar}(\vxi-\vxi')\wedge\x'\right)}.
\label{com-HU'}
\ee  
Inserting the semiclassical approximation \eref{SCUxi2} for $U'(\vxi,t)$  in
this equation, the full $4^L$-dimensional integral can then be evaluated by stationary phase.
First we note that the stationary conditions are just that $\vxi'=\vxi$ and that
$\x'=\x(\vxi',t)$, according to \eref{dSdxi}, i.e. the stationary centre 
that corresponds to the given chord is specified by the classical transformation. 
The Hessian determinant of the phase is unity,
so that, within the semiclassical approximation,
\be  
[\opH',\opU'(t)](\vxi) = \left[ H'(\J\frac{\der S'}{\der \vxi}-{\vxi\over 2})-
H'(\J\frac{\der S'}{\der \vxi}+{\vxi\over 2})\right]\>U'(\vxi,t).
\label{SC-com-HU'}
\ee

In the next section we will verify that the square brackets on the right specify
the time derivative of $S'(\vxi,t)$.
Though it is possible to derive directly the Hamilton-Jacobi equation 
that determines the evolution of the chord generating function
for the transported canonical transformation (\ref{classical}) following \cite{RiosOA},
the phase space geometry of fitting polygons is quite complicated. It is worthwhile
instead to invest on the double phase space approach followed by \cite{OsKon}. The great
advantage is that then chords and centres merely define conjugate planes in the 
enlarged space, so that their relation is entirely analogous to the conjugacy of
positions and momenta in ordinary phase space. In contrast, 
equation \eref{HJX2} appears strange indeed
when examined in ordinary phase space, if it is recalled that ordinary
Hamilton-Jacobi equations depend on only half of the phase space variables.

\section{Double phase space}

It might seem perverse to double the phase space of classical mechanics,
which is already a doubling of position space. Nonetheless, we are here
concerned with representing operators, commonly represented by both {\it bra} and
{\it ket} spaces, so it is not surprising that the classical correspondence
generally calls for the doubled classical space. Observables are deceptively simple in 
this respect, but, corresponding to unitary transformations, there arises 
an attractive and simple geometrical picture for the canonical transformations,
$\C:x_-\rightarrow x_+$, defined in the original phase space. Indeed the 
canonical property demands that all closed curves, $\gamma_-$,  be mapped
onto closed curves, $\gamma_+$, such that
\be
\oint_{\gamma_-} p_- \cdot \rmd q_- = \oint_{\gamma_+} p_+ \cdot \rmd q_+.
\label{canonical}
\ee  
Therefore, the definition of the double momentum space, $P=(-p_-, p_+)$,
and the double positions, $Q=(q_-,q_+)$, allows us to reinterpret the canonical 
condition as
\be
\oint_{\Gamma} P\cdot \rmd Q = 0,
\ee
where $\Gamma=(\gamma_-,\gamma_+)$. These are arbitrary closed curves on 
the $(2L)$-dimensional surface defined by the one-to-one function, $x_+=\C(x_-)$,  
within the $(4L)$-dimensional {\it double phase space} $X=(P,Q)$. 

In other words, the action, or symplectic area for any closed curve drawn 
on the surface that defines the canonical transformation, $\C$, in double phase
space is zero, i. e. canonical transformations are described by
{\it Lagrangian surfaces} in double phase space. This Lagrangian property
allows us to define locally a function,
\be
S_\C(Q)=\int_{Q_0}^Q P_\C(Q)\cdot \rmd Q,
\ee
which is independent of the path followed between $Q_0$ and $Q$.
In its turn, this {\it generating function} defines the given Lagrangian
surface by the equations
\begin{eqnarray}
\frac{\der S_\C}{\der Q}=P_\C(Q),\>\>
{\rm or}\>\>{\der S_\C\over{\der q_+}} =p_+,\>\>{\der S_\C\over{\der q-}}=-p_-,
\end{eqnarray}
that determine implicitly the canonical transformation \cite{Goldstein}.  

Though the mapping $\C:x_-\rightarrow x_+=\C(x_-)$ is necessarily
univalued, no such restriction results on the function $P_\C(Q)$, 
defined by the same $(2L)$-dimensional surface. We cannot define 
generating functions using $x_-$ or $x_+$ as independent variables,
because these do not lie on Lagrangian planes in the double phase space,
i. e. there is no constraint that either side of (\ref{canonical}) be
necessarily zero. What is allowed and often desirable is to apply
linear canonical transformations to the double phase space,
$X\rightarrow X'$, which leave invariant the Lagrangian property
for any surface, including $P'=0$. Then we may define
a new generating function in the new variables 
such that $P'_\C(Q')= \der S'_\C/ \der Q'$. 

All the commonly used generating functions \cite{Goldstein} are obtained
by the application of canonical $90^o$ rotations to the single phase spaces,
$q_{\pm}\rightarrow p_{\pm}$, separately or in combination. Obviously,
there exist unlimitted other possibilities in double phase space \cite{AmHu80},
but we will here be concerned only with the special canonical variables, 
\be
Q'=\x=\frac{x_+ + x_-}{2}, \>\>P'=\y= \J(x_+ -x_-)=\J\vxi.
\label{rotation}
\ee 
Instead of the previous $90^o$ rotations, this transformation to canonized
centre and chord variables is more like a $45^o$ rotation in double phase space.
The plane $\y=0$ (or $\vxi=0$) clearly specifies the identity transformation, $\vecI$,
which corresponds to a Lagrangian plane. Actually, all planes $\y=constant$ are
uniform translations, $T_\vxi$, by the vector $-\J\y=\vxi$, whereas each plane
defined by a constant $\x$ identifies the reflection $R_\x$. Unlike the Lagrangian 
plane $(q_-,q_+)$, the planes $\y=0$ and $\x=0$ can be considered 
as phase spaces on their own: the space of reflection centres 
(Weyl space) and the space of translation chords.
But it must be remembered that these are Lagrangian as far as the double 
phase space action, or symplectic form is concerned. 
Therefore, the mapping $\C$ defines implicitly the local
function $\y_\C(\x)$ in terms of the generating function $S_\C(\x)$:
$\y_\C(\x)=J\vxi_\C(\x)=\der S_\C /\der\x$, which provides a double phase space interpretation
for the relation between centres and chords in (\ref{dSdX}). 
Alternatively, the generating function $S_\C(\y)$ can be defined, 
such that $\x_\C(\y)=\der S_\C/\der\y$, corresponding to the chord generating function (\ref{dSdxi}).

Rather than derive the intricate single phase space geometry for
the evolution of the chord generating function, it is much simpler
to obtain the {\it Heisenberg form} of the Hamilton-Jacobi equation corresponding to
\eref{HJX2} from the double phase space dynamics. First, we note
that inserting \eref{dSdX} into \eref{HJX2} determines the double phase space
Hamiltonian as
\be 
\fl I\!\!H'(X)=I\!\!H'(\x,\y)=H'(\x-\J \y/2)-H'(\x+\J \y/2)=H'(x_+)-H'(x_-).
\label{Heisham}
\ee
This should be contrasted to the {\it Schroedinger} double phase space Hamiltonian
obtained from \eref{HJX}, which is simply
\be
I\!\!H(X)=I\!\!H(\x,\y)=H(\x-\J \y/2)=H(x_+).
\ee
In each case $I\!\!H(X)$ generates the Hamiltonian flow in double phase so as 
to evolve any Lagrangian surface acording to the respective 
Hamilton-Jacobi equation. The Heisenberg double Hamiltonian \eref{Heisham} 
determines the phase in Marinov's path integral for the Wigner function \cite{Marinov}. 

It must be stressed that the motion generated by the double Hamiltonian,
$I\!\!H'(X)$ in \eref{Heisham}, is purely classical, albeit in double phase space.
This is in general quite different from the Liouville flow in single phase space.
Generally, it is only within the invariant plane, $\y=0$, or $\vxi=0$, 
that the flow is Liouvillian, as will be shown in section 7. 

Generally, the double phase space motion generated by $I\!\!H'(\x,\y)$
in \eref{Heisham} depends on both the initial centre, $\x_0$ and
on $\y_0$. Indeed, double phase space can be considered as an incorporation
of the dynamics of pairs of orbits, starting from $x_{-0}$ and $x_{+0}$, 
into a  single Hamiltonian scheme. Thus, it is only within the 'horizontal' 
zero chord plane of double phase space
that the single and double phase space motions always coincide,
as will be further discussed in the next section. However, in the special case
of quadratic Hamiltonians, $H'(x)= -a\wedge x + x\cdot\mathbf B x$, where $\mathbf B$
is an orthogonal matrix, the motion of chords and centres becomes independent.
This follows simply from \eref{Heisham}, so that
\be
I\!\!H'(\x,\y)= a\cdot \y-2\x\cdot\mathbf B\J\y,
\label{smally}
\ee
and Hamilton's equations for the conjugate variables $\x$ and $\y$ become 
\be
\dot\x=-\frac{\der I\!\!H'}{\der\y}= -a-2\J\mathbf B\x
\ee
and
\be
\dot\y=\frac{\der I\!\!H'}{\der\x}=-2\mathbf B\J\y.
\ee
In the case of the centres, this is exactly the same as $\dot x$ in single phase space
and merely reflects the fact that the centre between two points satisfies the same linear
equation as each of them individually. This fact is the basis of the classical propagation
of the Wigner function and Weyl symbols for quadratic Hamiltonians. What is not so familiar
is that the chord function and chord symbols also propagate classically \cite{BroAlm04}. 
In the case of a homogeneous quadratic form $(a=0)$, Hamilton's
equations for $\dot\vxi=\J\dot\y$ and for $\dot x$ again coincide according to \cite{BroAlm04}, 
but the chords are insensitive to the linear part $a\wedge x$. 
The latter generates a translation, which appears in the chord symbol as the phase factor in \eref{shift}.

Coordenatizing locally the moving surface driven by $I\!\!H'(X)$ as
$\y=\y'(\x,t)$, the solution is the generating function
\be
S'(\x,t)=\int_{\x_0}^\x \y'(\x,t)\cdot\rm d\x.
\ee
The conjugate action is defined via the Legendre transform,
\be
S'(\y,t)=\x'\cdot\y-S'(\x',t),
\label{Legendre}
\ee
in the usual way, i. e. $\x'(\y,t)$ is specified by the condition that
\eref{Legendre} is stationary with respect to $\x'$. Taking now the 
full derivative of \eref{Legendre} with respect to the time and using
the fact that $\der S'(\y,t)/\der\y=\x'(\y,t)$, then leads to
\be
\frac{\der S'(\y,t)}{\der t}=-\>\frac{\der S'(\x',t)}{\der t}=I\!\!H'(\x'(\y,t),\y).
\label{dert}
\ee
Reintroducing \eref{dSdxi} into \eref{dert} and again recalling that $\vxi=-\J y$,
leads to the chord evolution equation corresponding to the Heisenberg evolution as
\be
\frac{\der S'(\vxi,t)}{\der t}-H'(\J\frac{\der S'}{\der \vxi}+{\vxi\over 2})+
H'(\J\frac{\der S'}{\der \vxi}-{\vxi\over 2})=0.
\label{HJ-xi-Heis}
\ee
Of course, we can also obtain the partial differential 
Hamilton-Jacobi equation for the Schroedinger
action $S_t(\x)$ by exactly the same procedure, so that
\be
\frac{\der S_t(\vxi)}{\der t}-H(\J\frac{\der S_t}{\der \vxi}+{\vxi\over 2})=0.
\ee

\section{Operator evolution revisited}

To apply the above theory to the evolution of quantum operators it should be recalled
that they form a Hilbert space of Hilbert-Schmidt operators 
with the {\it scalar product} \cite{Vor76,Littlejohn95},
\be
\langle\!\langle A|B\rangle\!\rangle= {\rm tr}\> \opA^{\dagger}\widehat B,
\label{scalarprod}
\ee
defined in terms of the adjoint operator, $\opA^{\dagger}$. 
When the trace is defined, the Heisenberg (or von Neumann) evolution 
can be considered as the action of a {\it unitary super-operator}, 
since it preserves the scalar product. General super-operators always preserve
the trace of the self-adjoint density operator, ${\rm tr\oprho}$, 
but not necessarily ${\rm tr{\oprho}^2}$ \cite{Kraus}. 

Each foliation of double phase space by parallel Lagrangian planes corresponds
to a possible operator representation. 
Perhaps the most common representation relies on the position projection operators,
$\langle\!\langle Q|=|q_-\rangle\langle q_+|$, so that
\be
\langle\!\langle Q|A\rangle\!\rangle=\langle q_+|\opA|q_-\rangle=
{\rm tr}\>|q_-\rangle\langle q_+|\opA,
\ee
where the Lagrangian planes are just $Q=(q_-,q_+) =constant$. From this one can switch to momentum, 
or various mixed representations through Fourier transformations, corresponding to $90^o$
rotations in double phase space. The Weyl representation, based on the self-adjoint operator,
$\opR_\x$, then corresponds to the double phase space rotation
\label{rotation}, so that for $Q'=\x$,
\be
\langle\!\langle Q'|A\rangle\!\rangle=2^L {\rm tr}\> \opR_{\x}\opA=A(\x),
\ee
and the Lagrangian basis in double phase space are the reflection planes, $\x=constant$.
The Fourier transformation \eref{quantreflection} then brings in the translation operator,
whose adjoint is $\opT_{-\vxi}$. This is represented in double phase space 
by the new Lagrangian plane, $P'=\y=\J\vxi=constant$, so that
\be
\langle\!\langle P'|A\rangle\!\rangle= {\rm tr}\> \opT_{-\vxi}\opA=A(\vxi).
\ee
In each case the representation in terms of a set of Lagrangian planes, $Q'$, is complementary
to the conjugate representation in terms of $P'$, which is obtained by a Fourier transform.
Thus, Heisenberg's uncertainty principle manifests itself in double phase space.
Further discussion is presented in  reference \cite{Chountasis}.

The suggestive use of a Dirac notation for operators, in direct analogy to that
commonly reserved for states in Hilbert space, is supported by the manner in which
the evolution of unitary operators is semiclassically related to the evolution of
Lagrangian surfaces, in double phase space. In the case of the chord propagator,
combining \eref{HJ-xi-Heis} with \eref{SC-com-HU'} we verify that
\be
[\opH',\opU'(t)](\vxi) = {\hbar\over \rmi}\frac{\der U'}{\der t}(\vxi,t),
\ee
to lowest order in Planck's constant. In the following sections we study
the evolution of the particular unitary operators, translations and reflections,
which are the basis of the present theory.
 
The preceding theory is concerned with the evolution of unitary operators
that always correspond to Lagrangian surfaces in double phase space.
The next section focuses on evolved unitary translation operators which allow us
to propagate arbitrary operators through their chord symbols. But first,
it should be noted that the present semiclassical theory can also be applied
directly to other operators that also correspond classically to Lagrangian
surfaces in double phase space. 

The most important case is that of projectors,
or pure state density operators, $|\psi\rangle\langle\psi|$. Indeed, it is quite
usual for the semiclassical description of a state $|\psi\rangle$ to be based on
a Lagrangian surface in single phase space, a torus for a bounded state, e.g. a closed curve
in the case that $L=1$. Then the projector onto this state is supported by
the product manifold with doubled dimension, e.g. a two-dimensional torus
if $L=1$. Another interesting case is that of a dyadic operator,
$|\psi\rangle\langle\psi'|$, which is useful to describe transitions. 
In the case that $|\psi\rangle$ and $|\psi'\rangle$
are eigenstates of the same observable, the Weyl representation of this
operator is known as a Moyal fuction, or cross-Wigner function \cite{Moyal, Dodonov86}. 
Its  semiclassical form in single phase space has been previously studied \cite{Alm84}, 
relying on chords (and their centres)
that connect the pair of distinct single tori,
but the double phase space picture is simpler because a single Lagrangian surface
is involved as in the case of the projectors.

In all these cases, caustics cannot generally be avoided.
Taking the product of the torus in $x_-$
with the torus in $x_+$, results in a surface with the same dimension as 
the single phase space in which each torus is embedded, but which projects down singularly 
as the single torus into either of $x_\pm$ . But, as we have seen, $x_\pm$ are not 
Lagrangian surfaces in double phase space with which to view caustics of
the double torus in the sense that they cannot supply the variables for a generating function.
The $\vxi=0$ plane within double phase space, in which the
Wigner function is defined, is Lagrangian and the caustic that arises along
the torus in the single space picture \cite{Ber77} can now be reinterpreted 
as the fold caustic for the projection of the double
torus onto this particular coordinate plane. The section of the double torus
with the plane $\vxi=0$ is identical to the single space torus, 
just as the sections with $x_-$ and $x_+$.
If $L>1$, the Wigner caustic has a higher dimension than the single torus, 
though the latter is included as a higher singularity \cite{AlmHan82}. In all cases, the
Wigner caustic results from the projection of the double torus onto the $\x$ plane.
Further application of semiclassical methods in double phase space 
are presented in \cite{Littlejohn95}.

The projection of the double torus onto the chord plane, $\x=0$, is even more singular.
Indeed, the origin of this plane, $\vxi=0$, is the image of the entire single torus
The approximation of the chord function in this region is studied in \cite{AlmVal04},
as well as the caustic for maximal chords of the single torus.
In any case, it is clear that a simple Fourier switch between the Weyl 
and the chord representations does not deliver us from caustics 
and the problem of their evolution. The alternative that is pursued in sections 7
and 8 is to propagate either the translation, or the reflection basis operators.

\section{Propagating translations}

It was reviewed in section 2 that the chord representation may
be considered as the decomposition of an arbitrary operator into a superposition
of unitary translations. Hence, the general evolution
of operators in this representation can be reduced to the propagation of this 
special class of operators, each of which corresponds to a basis 
plane $\y=constant$ in double phase space. These can be  pictured as horizontal, 
if we interpret $\y$ as the double momentum, $P'$, and the $\y=\vxi=0$ plane 
corresponds to the identity transformation. 
Each of its points results from the intersection with a vertical, $Q'=\x=constant$,
plane that corresponds to a reflection, being that the plane $\x=0$ corresponds to the
operator for reflection through the origin of ordinary single phase space, 
which is also known as the parity operator.
 
It is immediately obvious from the form of the double Hamiltonian
for Heisenberg propagation \eref{Heisham} that the $\y=0$ plane is always invariant.
This merely expresses the fact that the corresponding canonical transformation 
remains the identity for all time, just as  the quantum identity operator, 
$\opI$, is invariant  for arbitrary Heisenberg 
propagation. Therefore, classical-quantum 
correspondence becomes exact in the limit of small chords. This confirms the interpretation
in \cite{AlmVal04} of the long chords as responsible for the essentially quantum 
long range correlations of the density operator. Within the purely quantum domain,
the combination of the unitarity of the Heisenberg super-operators, which preserves 
the trace of products of operators \eref{scalarprod}, with the invariance of $\opI$, 
results in the invariance of the trace itself. This property must be demanded of
the nonunitary super-operators for open systems, 
which are no longer generated by the Heisenberg equation\cite{Kraus}.

The flow within the invariant classical plane
results from the expansion of the double phase space Hamiltonian, 
$I\!\!H'(\x,\y)$ in \eref{Heisham}, for a fixed centre $\x$,
\be
I\!\!H'(\x,\y)=\frac{\der H'}{\der\x}(\x,0)\wedge\y + \GO(\y^3).
\label{smallH}
\ee
Hence, Hamilton's equations in double phase space for the centres, $\x$, within 
the classical invariant plane are identical to the equations of motion generated
by the single phase space Hamiltonian, $H'(x)$.   

General translations, $\C=T_\vxi$, are not invariant for arbitrary Heisenberg propagation.
The initial Lagrangian surface that will be evolved by $\C'(t)=T'_\vxi(t)$
is a horizontal plane, so that analogy with single phase space indicates that the chord
symbol is
\be
T_{\vxi}(\vxi')=\delta(\vxi-\vxi'),
\label{delta'}
\ee
whereas the Weyl representation is the plane wave \cite{Alm98},
\be
T_{\vxi}(\x)= \exp {\left(-{i\over \hbar} \x \wedge \vxi\right)}.
\label{planewave'}
\ee
The chord symbol is not in a simple semiclassical form, because the horizontal
planes project singularly onto the vertical axis, but \eref{planewave'} is in the
form \eref{SCUX} with the classical action $S_\vxi(\x)=-\x \wedge \vxi$ and the
tangent matrix $\M_\vxi$, defined by \eref{linearize} is twice the unit matrix. 
Inserting this linear form 
of the generating function in \eref{dSdX} results in the same chord $\vxi$
being placed on all the centres $\x$. This need no longer hold after a
classical Heisenberg evolution,
so that the evolved actions $S'_\vxi(\x',t)$ obtained from the Hamilton-Jacobi
equation \eref{HJX2} will generally have second and higher order terms and
$\M_\vxi(\x',t)$ will no longer be proportional to the unit matrix. Unless the evolution
proceeds to the possible production of a vertical fold, i.e. a centre caustic,
the phase $\theta$ in \eref{SCUX} will be the same as in \eref{planewave'}.
Thus the general semiclassical form of the distorted translation 
in the centre representation is
\be
T'_\vxi(\x',t) = 2^L |\det (1+{\M'_\vxi(\x',t)})|^{-\frac{1}{2}} \exp [i\hbar^{-1}S'_\vxi(\x',t)].
\label{evoltrans}
\ee
The identification of the Hermitian conjugate of $\opT'_\vxi(t)$ with $\opT'_{-\vxi}(t)$,
i.e. the inverse operator, establishes that $S'_{-\vxi}(\x,t)=-S'_\vxi(\x,t)$.

An important constraint on the classical Heisenberg evolution of the plane
that describes a finite translation is that it cannot touch, or intersect the invariant identity 
plane, $\y=0$. This means that the distorted translation never develops a fixed point.
Indeed, the number of fixed points is invariant throughout the evolution. Furthermore,
the linearization of the monodromy matrix near each fixed point is also invariant.
Thus, summing over the fixed points we obtain a semiclassical invariance of Tabor's
version \cite{Tabor} of Gutzwiller's trace formula \cite{Gutzwiller}, 
once the quantum invariance of the trace of any operator is recalled. 

If we now adopt the Weyl representation for the Heisenberg
evolution $\opA'(t)$ of an arbitrary operator $\opA$, \eref{chordrep} leads to 
\be
A'(\x',t)= \int \frac{\rmd\vxi}{(2\pi\hbar)^L} \>A(\vxi) \> T'_{\vxi}(\x',t).
\label{chord-centre}
\ee
Hence, the distorted translation, $T'_{\vxi}(\x',t)$, is identified as 
the mixed {\it chord-centre propagator}. In view of the previous discussion,
the simple semiclassical approximation \eref{evoltrans} then supplies an initially
caustic free approximation for the Weyl symbol (or Wigner function) 
that evolves from a given chord symbol. 
This is not the case of the {\it chord-chord propagator} 
that follows from the evolution of the chord symbol of $\opA'(t)$ given by \eref{chordrep}: 
\be
A'(\vxi',t)= \int \frac{\rmd\vxi}{(2\pi\hbar)^L} \>A(\vxi) \> T'_{\vxi}(\vxi',t).
\label{chord-chord}
\ee
Even though the invariance of ${\rm tr} \opA'(t)$, implies that $A'(0,t)=A'(0,0)$ for all $t$,
this propagator evolves from the $\delta$-function \eref{delta'} into
a semiclassical form that must negotiate caustics through Airy functions,
or other higher {\it diffraction catastrophes} \cite{Berry-Up, livro}. Using the
product rules for chord symbols \cite{Alm98} it is possible to express the chord-chord
propagator directly in terms of the chord symbols for the elementary evolution operators
$\widehat V'_t$ in the full Heisenberg propagation \eref{quantum}:
\be
T'_{\vxi}(\vxi',t)= \int\! \rmd\eta \>\>\> V'_t\left(\eta+\frac{\vxi'-\vxi}{2}\right)\>
 V'_t\left(\eta-\frac{\vxi'-\vxi}{2}\right)^*
\exp\left({i\over\hbar}\eta \wedge\frac{\vxi'+\vxi}{2}\right).
\label{exact-chch}
\ee
It is interesting to note that this exact expression has the same form as the
transformation that defines the chord function $\chi(\vxi)$ for a pure state density operator
from the wave function $\langle q|\psi\rangle$ \cite{AlmVal04}. Stationary phase evaluation of
\eref{exact-chch} far from caustics leads to a superposition of terms of the general semiclassical
form \eref{SCUxi2}. 

As noted at the end of section 6, we are dealing with a general semiclassical theory 
for the evolution of operators, each of which corresponds to a Lagrangian surface in double
phase space. The translation operators are an important particular case, but
if the operator $\opA$ that is propagated is related to a particular Lagrangian
surface of its own, one may as well evolve it directly in the Weyl representation
using \eref{SCUX}, instead of integrating the evolved basis operators. Indeed, far from
a caustic, the stationary phase evaluation of \eref{chord-centre}, or \eref{chord-chord}
will give the same result as \eref{SCUX}. However, specially for pure state density operators,
the whole region of small chords, that is identified with the near-classical part
of the density operator \cite{AlmVal04}, lies in the neighbourhood of a caustic of the chord function
as well as a caustic of the Wigner function. Furthermore, the phenomenon of {\it decoherence},
i.e. the evolution of the density operator in an open system (into a mixed state) \cite{Giulini},
leads to a cut-off for the contribution of the longer chords \cite{BroAlm04}. 
It is thus important to determine the semiclassical evolution of the neighbourhood
of the invariant plane $\y=0$ (or $\vxi=0$) in double phase space. 
The difficulty is that it is necessary to keep track of the (small) chords throughout
the evolution, whereas the chord-centre propagator in this section depends on the initial chord,
but is later only described by the centres. 
The situation is clarified in the following section by the alternative definition of the mixed
centre-chord propagator.

\section{Propagating reflections}

The Heisenberg evolution of arbitrary operators, $\opA$, according to \eref{quantum},
is given by \eref{Weylrep} as a superposition of evolved reflection operators. 
(Though density operators propagate backwards in time.) Taking the Weyl symbol
of the evolved operator $\opA'(t)$, then leads to
\be
A'(\x',t)= \int d\x \>A(\x) \> 2^L R'_{\x}(\x',t),
\label{Wsymbolprop}
\ee
where we introduce in the integrand the Weyl symbol for $\widehat R'_\x(t)$, 
the quantum reflection through $\x$ seen in a moving coordinate frame. 
Thus, \eref{Wsymbolprop}  identifies $2^L R'_{\x}(\x',t)$ 
as the propagator for Weyl symbols and Wigner functions 
(with backward propagation in the latter case). 
This propagator can be expressed exactly as a path integral \cite{Marinov} and can also be obtained 
exactly \cite{RiosOA} from the Weyl symbols of the operators $\widehat V'_t$ that are 
responsible for the Heisenberg evolution \eref{quantum}, 
using the Weyl product rules \cite{Moyal}:
\be
\fl 2^L R'_{\x}(\x',t)= \int\! \rmd x \>\>\> V'_t\left(\frac{\x'+\x}{2}+x\right)\>
 V'_t\left(\frac{\x'+\x}{2}-x\right)^*
\exp\left({2i\over\hbar}x \wedge(\x'-\x)\right).
\label{exact-cece}
\ee
This formula is the companion of \eref{exact-chch} and it has the same structure
as the transform that defines the Wigner function for a pure state
in terms of the corresponding wave function \cite{Wigner}.

Initially, for
$t=0$, \eref{Wsymbolprop} reduces to the known result that
\be
R_{\x}(\x')=2^{-L}\delta(\x-\x'),
\ee
i.e. the Weyl symbol for the reflection through the point $\x$ is just a
$\delta$-function \cite{Grossmann}. This is analogous to the
fact that position states in single phase space are represented by $\delta$-functions 
in the position representation, because we associate
reflection centres to positions, $Q'=\x$, and canonized translation chords, $P'=\y=\J\vxi$,
to momenta. Thus, it is no surprise that in 
the conjugate chord representation we have
\be
R_{\x}(\vxi)= \exp {\left({i\over \hbar} \x \wedge \vxi\right)},
\label{planewave}
\ee
i.e. the reflection operator is represented by a plane wave \cite{Alm98}. 
Note that the double phase space 
relation between reflection operators, viewed in the translation (chord)
representation \eref{planewave}, and translation operators in the Weyl
(reflection centre) representation is in perfect analogy to that between
the relation for position eigenstates viewed in the momentum representation
and momentum eigenstates in the position representation. 

In the special case where the Hamiltonian, $\opH'$, is quadratic, 
so that the corresponding classical transformations are linear,
the propagator for Weyl symbols remains a Liouvillian $\delta$-function for all time,
\be
R'_{\x}(\x',t)=R_{\x(t)}(\x')=2^{-L}\delta(\x(t)-\x'),
\label{deltaprop}
\ee
where $\x(t)$ is the classical trajectory for $\x$. 
The double phase space point of view shows how special this situation is:
it is a canonical transformation that takes vertical planes into vertical planes.
The Appendix discusses how
any nonlinearity immediately introduces caustics into this centre representation,
because the transformed surface representing  the canonical transformation, 
$R'_\x(t)$, must have a vertical tangent at $\y=0$, as shown in Fig.3(b).
The nonlinear distortion of a reflection in the original single phase space
is sketched in Fig.3(a).

In contrast, using the chord representation for these propagated reflections, $\opR'(t)$,
the Heisenberg evolution merely distorts the plane wave \eref{planewave}, which can only 
develop a caustic after a finite time. Even so, the invariance of the number of
fixed points for classical Heisenberg evolution commented in the previous section
prevents the evolved vertical plane in double phase space from intersecting, or even
touching the invariant plane $\y=0$ a second time.
Comparing \eref{planewave} with \eref{SCUxi2}
shows that the exact $R'_\x(\vxi',0)$ is already in its semiclassical form,
with $S'_{\x}(\vxi',0)=\x \wedge \vxi'$. The classical Heisenberg evolution 
\eref{classical}, i.e. the Hamilton-Jacobi equation \eref{HJ-xi-Heis}, 
takes this into 
\be
S'_{\x}(\vxi',t)=\x(t)\wedge\vxi'+\GO(\vxi'^3), 
\label{smallchord-S}
\ee
which is still
an approximate linear reflection about $\x(t)$ for small chords. All even powers in the 
components of $\vxi$ are missing in $S'_{\x}(\vxi',t)$, because the involution property
of the reflections, that is $R_{\x}^2=\vecI$, the identity, 
remains invariant for coordinate changes; hence $\x(-\vxi,t)=\x(\vxi,t))$.
\begin{figure}
\includegraphics[width=6.5cm]{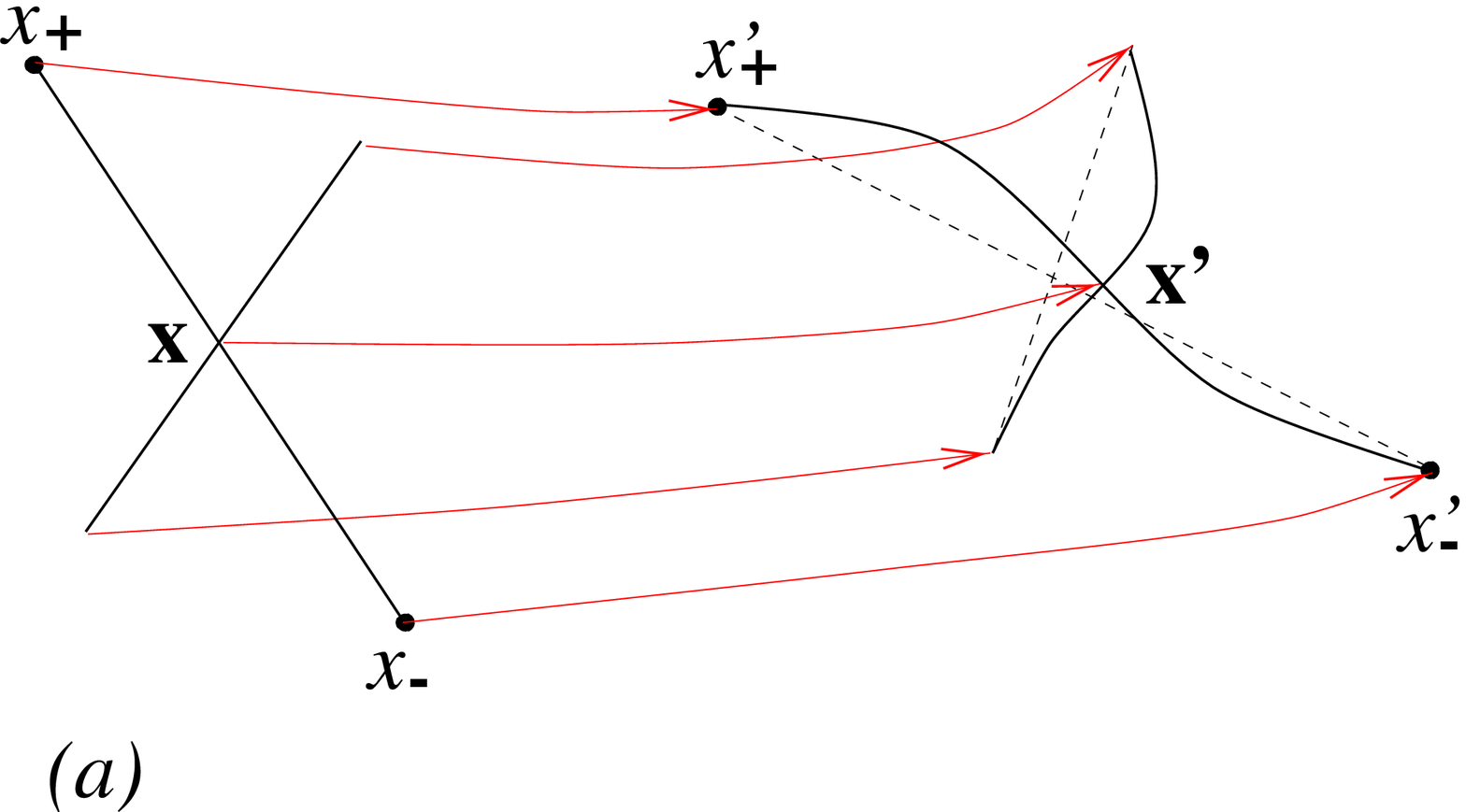}
\includegraphics[width=8cm]{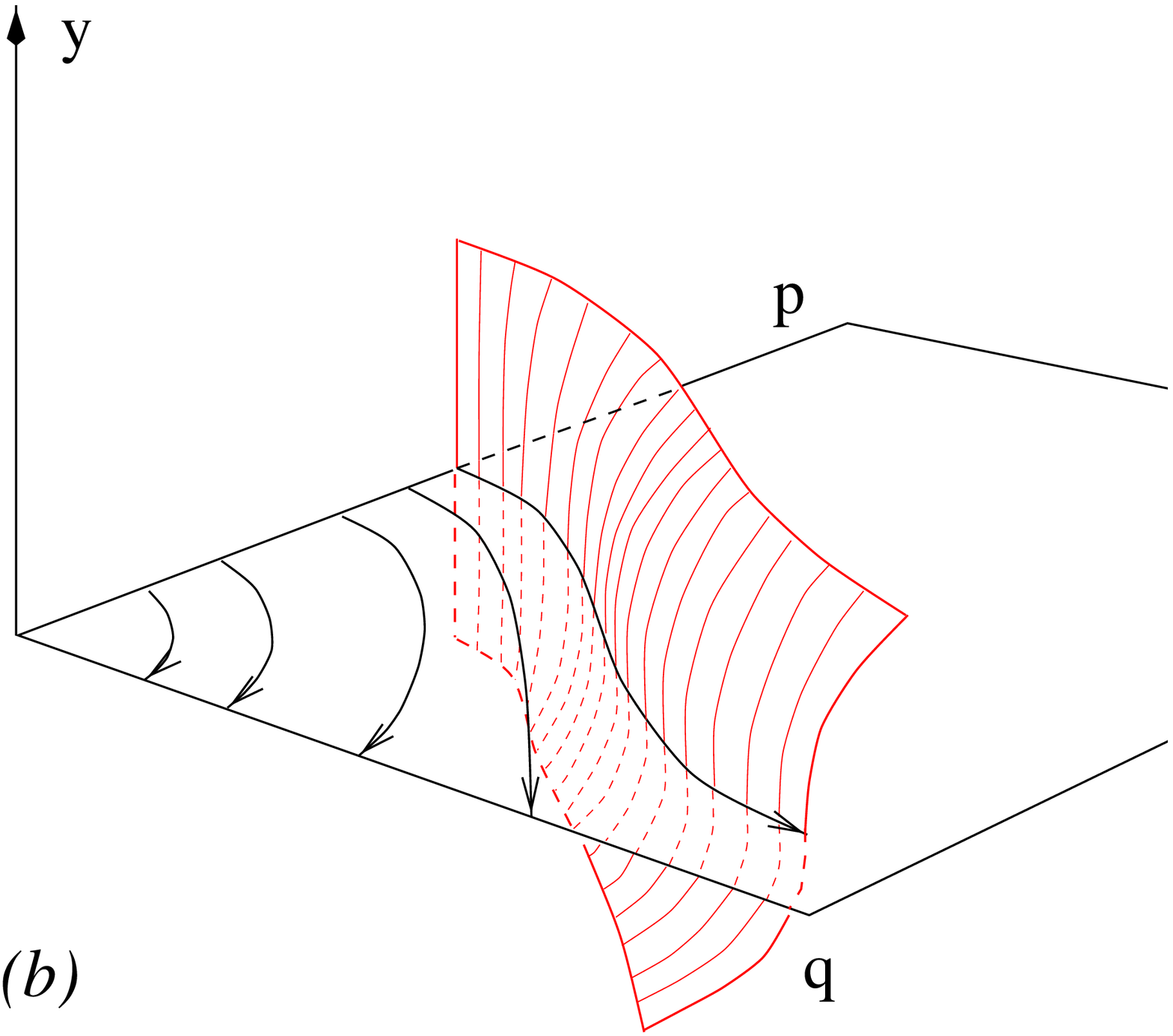}
\caption{(a) A nonlinear classical Heisenberg evolution evolution maps the original reflection centre,
$\x\rightarrow\x'(t)$, but the linear reflection about this point is only a local linear approximation 
to the evolved transformation, $R'_\x(t)$: 
The centres, $\x'(\vxi')$, of finite chords do not generally coincide with $\x'(t)$. 
However, for all chords, $\x'(\vxi')=\x'(-\vxi')$. (b) In double phase space, a linear reflection
is represented by a vertical plane. This is curved by a nonlinear evolution, 
but the tangent is still vertical at the identity plane, $\y=0$.}
\label{fig3}
\end{figure}

There is a simple semiclassical form for the mixed {\it centre-chord propagator} 
that defines the chord symbol $A'(\vxi',t)$ for the operator
$\widehat A'(t)$, which has evolved according to the Heisenberg equation from
the operator $\widehat A$ with Weyl symbol $A(\x)$:
\be
A'(\vxi',t)=\int \rmd\x \> A(\x)\> 2^L R'_{\x}(\vxi',t).
\label{centre-chord}
\ee
The mixed propagator has the semiclassical form defined by a single
evolving classical generating function, $S'_{\x}(\vxi',t)$, which is a special case
of \eref{SCUxi2}:
\be
R'_{\x}(\vxi',t) = |\det (1-\M'_{\x}(\vxi',t))|^{-\frac{1}{2}}  \exp [i\hbar^{-1}S'_{\x}(\vxi',t)],
\label{centre-chord-SC}
\ee
where the monodromy matrix $\M'$ for the transformation from $\x-\vxi/2$ to $\x+\vxi/2$
is obtained from the evolving action in \eref{Cayley'}. The familly resemblance
between the semiclassical expressions \eref{centre-chord-SC} for $R'_\x(\vxi',t)$ and
\eref{evoltrans} for $T'_\vxi(\x',t)$ can be pushed all the way by the identity,
\be
T'_\vxi(\x',t)= {\rm tr}\>\opT'_\vxi(t) 2^L\opR_{\x'}={\rm tr}\>2^L\opR{_\x'}(-t)\opT_\vxi=R'_{\x'}(\vxi,-t),
\label{propidentity}
\ee 
where \eref{quantum} is used for $\opT'_\vxi(t)$ and $\opR_\x'(-t)$. It follows that 
$S'_{\x}(\vxi',t)= S'_{\vxi'}(\x, -t)$ and hence both semiclassical propagators will be
free of caustics for the same interval of time. The identity \eref{propidentity} expresses
the fact that Heisenberg evolution defines unitary super-operators, so that a complex conjugate
kernel is obtained by reversing the time. This unitarity is also apparent in \eref{exact-cece}.
  
In the limit of small chords the amplitude in \eref{centre-chord-SC} tends to one, 
just as in \eref{planewave}. In fact, it is now legitimate to investigate this limit 
and thus keep only the lowest term in \eref{smallchord-S}, since $\vxi'$ is the free variable.
In this region even the distorted reflexion operator takes the simple form 
\be
R'_{\x}(\vxi',t)= \exp {\left({i\over \hbar} \x(t) \wedge \vxi'\right)}.
\label{planewave-t}
\ee
Here, $\x(t)$ is the Liouville orbit of $\x(0)=\x$ in single phase space, because
we are restricting the double phase space motion to the invariant plane $\y=0$, where
the double Hamiltonian is just \eref{smallH}.
Thus, the mixed centre-chord propagation \eref{centre-chord} becomes approximately
\be
A'(\vxi',t)=\int \rmd\x \> A(\x)\> 2^L \exp {\left({i\over \hbar} \x(t) \wedge \vxi'\right)}.
\label{centre-chord-small}
\ee
The small chord approximation is thus a moving Fourier transform of the initial Weyl symbol.
At first sight, this may appear to be equivalent to generalizing the $\delta$-function 
centre-centre propagator \eref{deltaprop} for nonquadratic Hamiltonians, but this is not so.
Such a stronger approximation is equivalent to extrapolating \eref{centre-chord-small}
for all chords, which is not necessary in the chord representation.

Consider, for instance, the evolution of the pure state density operators studied in
\cite{AlmVal04}. Even though such operators may be associated 
to a double phase space Lagrangian surface, 
given locally by the functions $\vxi=\vxi(\x)$, or $\x=\x(\vxi)$, both the chord
and the centre representations have caustics in the part of this surface for which
$\vxi$ is small. Reference \cite{AlmVal04} developed a small chord approximation
which could be extended so as to overlap with the semiclassical approximation 
valid for long chords. Now we find that these different regions can be propagated separately.
For the long chords that carry information concerning long range quantum correlations, 
the simplest is to procede with a direct semiclassical evolution, 
based on the classical propagation of the
generating function, $S'(\vxi,t)$, for the Lagrangian surface corresponding to the density operator. 
As for the small chords in the near classical region, 
the best alternative is to use a chord basis, 
so that the evolution is pictured as a moving Fourier transform
from the initial Wigner function  \eref{centre-chord-small}.

\subsection{A simple cubic case}

The case of a quadratic Hamiltonian is exactly solvable, as mentioned previously. 
An exact result in the nonquadratic case is also available for the Hamiltonian 
\be
H'(x)=ap^3.
\label{p3}
\ee
We can indeed make the full development of the double phase space
Heisenberg Hamiltonian \eref{Heisham}, 
\be
I\!\!H'(X)=H'(\x+\vxi/2)-H(\x-\vxi/2)= 
a\left(3p^2\vxi_p+\frac{1}{4}\vxi_p^3\right).
\ee
However, it must be remembered that the relevant classical motion
occurs in the double phase space $X=(\x,\> \y=\J\vxi)$, so that the
conjugate pair of variables are $(\x_p=p, \y_p=-\vxi_q)$ and $(\x_q=q, \y_q=\vxi_p)$.
Thus, Hamilton's equations for $I\!\!H'(X)$ result in constant values for
$p$ and $\vxi_p$, whereas
\be
q(\vxi, t) = q + 3a(p^2-\frac{1}{4}\vxi_p^2)t
\label{qxit-cubic}
\ee
and $\vxi_q(t)=\vxi_q+6ap\vxi_p t$. Note that the motion within the invariant plane,
$\vxi=0$, is just the classical motion in single phase space:
\be
q(t) = q + 3ap^2t.
\label{qt-cubic}
\ee

In this simple case, we can verify that the chord symbol 
for the reflection through the point $\x$,
\be
R'_\x(\vxi',t)=\exp{\left(\frac{i}{\hbar}(p\vxi'_q-q(t)\vxi'_p+{at\over 4}\vxi_p^3)\right)}, 
\label{refl-p3}
\ee
satisfies
\be
\frac{\der R'_{\x}}{\der t}(\vxi',t) = - \frac{i}{\hbar}a\Biggl[
3p^2\vxi'_p+\frac{1}{4}{\vxi'_p}^3 \Biggr]
R'_{\x}(\vxi',t) 
\ee
by performing the integrations in the exact expression for the commutator \eref{com-HU'}.
Furthermore, the phase
\be
S'_\x(\vxi',t)=p\vxi'_q-q(t)\vxi'_p+{at\over 4}{\vxi'_p}^3
\ee
in \eref{refl-p3} satisfies the Hamilton-Jacobi equation
\eref{HJ-xi-Heis}, so that the semiclassical form of the mixed propagator
\eref{centre-chord-SC} is exact in this example, given that the amplitude is constant. 
Nonetheless, the motion $\x'(\vxi',t)$
obtained from the derivative of $S'_\x(\vxi',t)$ according to \eref{dSdxi}, 
i.e. \eref{qxit-cubic}, is not Liouvillian unless $\vxi'=0$.

Inserting \eref{refl-p3} into the general evolution formula \eref{centre-chord}
results in
\be
A'(\vxi',t)=\int \rmd\x \> A(\x)\> 2 
\exp\left[\frac{i}{\hbar}\Biggl(p\vxi'_q-q(t)\vxi'_p+{at\over 4}{\vxi'_p}^3\Biggr)\right] .
\label{centre-chord-p3}
\ee
Thus, the Fourier kernel that initially transforms from the Weyl representation to
the chord symbol becomes non-Liouvillian, as well as nonlinear, and the cubic term in the
phase cannot generaly be neglected. 

Let us now consider the propagation of the Wigner function for the
position projector $|q_0\rangle\langle q_0|$, which is just $W_0(\x)=\delta(q-q_0)$.
This corresponds classically to the evolution of the vertical straight line
which is bent into a parabola under the action of the Hamiltonian \eref{p3}.
According to \eref{centre-chord-p3} the evolving chord function will be
\be
\chi'_{0}(\vxi',t)=\int \rmd p\rmd q \> \delta(q-q_0)\> 2 
\exp \left[{i\over\hbar}\Biggl(p\vxi'_q-q(t)\vxi'_p+{at\over 4}{\vxi'_p}^3\right)\Biggr],
\ee
which can be immediately integrated to yield
\be
\chi'_{0}(\vxi',t)=\left[\frac{2\pi\hbar}{3at\vxi'_p}\right]^{1/2}
\exp \left[{i\over\hbar}\left(-q_0\vxi'_p + {at\over 4}{\vxi'_p}^3 + 
\frac{{\vxi'_q}^2}{12at\vxi'_p}\right)\right].
\label{parabola}
\ee
This is the exact chord function for the parabola, though it is in
its semiclassical form \cite{AlmVal04}, so the exponent is just the chord
action. Unlike the corresponding Wigner function, there is no interference,
because a parabola  translated by $\vxi'$ intersects the original parabola
at a single point. In contrast, a parabola  reflected through $\x'$ intersects
twice, if $\x'$ is in the concave region, or not at all. The corresponding 
Wigner function, $W'_{0}(\x',t)$, is an Airy function, as can be verified by 
directly integrating the Fourier transform that relates these representations, 
using \eref{parabola}. The parabola itself is the Wigner caustic,
whereas the chord function has a nongeneric caustic for $\vxi'_p=0$, 
which corresponds to the translated parabola approaching the original 
parabola at infinity, for any horizontal translation. 

The small chord approximation \eref{centre-chord-small} in this case
merely misses the $(at/4){\vxi'_p}^3$ part in the exact expression \eref{parabola} 
for $\chi'_{0}(\vxi',t)$. However, it is a good approximation for small chords
even along the chord caustic. It is only if this locally valid expression is extrapolated
for all chords that its Fourier transform yields the crude Liouvillian approximation to
the Airy function evolution:
\be
W'_{0}(\x',t)=\delta(q'-q_0-3a{p'}^2t).
\ee


\section{Conclusion}

Previous semiclassical approximations for the evolution of the Wigner function,
or other oscillatory Weyl symbols have not dealt with the near-classical region where
the conjugate chords are small. However, this will be the dominant region in the
integral that determines the expected value of any observable, $\opA$:
\be
<\opA>= \int \rmd\x \>W(\x) \>A(\x).
\ee
We have shown that the simplest way to deal with the propagation in the near-classical
region is to resort to the mixed propagators between the chord symbol, $A(\vxi)$,
and the centre, or Weyl representation. These propagation kernels 
are defined by the Heisenberg evolution of different operators,
$2^L R'_\x(\vxi',t)$ and $T'_\vxi(\x',t)$,
namely reflections and translations, but they are identified by the relation \eref{propidentity}.

Though it is possible to dispense with the construction of a double space, 
the conceptual clarification also simplifies the calculation. From this
point of view, the transition between centres and chords lies in strict analogy
to that which relates the conjugate positions and momenta in ordinary phase space.
From this point of view, we have merely applied the general Maslov method 
for negociating caustics \cite{Maslov} to double phase space. It turns out
that caustics in the evolution kernel are not avoided by viewing the entire evolution
in neither the Weyl, nor the chord representation on their own, but we are guaranteed a simple semiclassical
form by alternating between them. Of course, the more intricate chord-chord, or centre-centre
propagations lie a mere Fourier transform away.

Whether considered in double, or in single phase space, the propagation of unitary
operators, such as the translations, or reflections which are the bases of the chord
and the Weyl representations, have a clear classical analog. This {\it classical
Heisenberg evolution} continuously distorts a fixed canonical transformation through
an evolving change of coordinates in phase space. Viewed in double phase space, this
is an evolving Lagrangian surface. Once this quantum-classical correspondence is 
established, we are free to also propagate other Lagrangian surfaces in the doubled space,
which, for instance correspond to projectors, or dyadic transition operators. In all cases,
the motion is driven in double phase space by a classical Heisenberg Hamiltonian, 
simply related to the single Hamiltonian by  \eref{Heisham}. Each trajectory in
double phase space corresponds to a pair of trajectories in single space,
which emmanate from the pair of points $x_\pm=\x\pm\vxi/2$. Only in the limiting case
where $\vxi\rightarrow 0$ does the motion of the centre $\x(t)$ coincide
with the single phase space trajectories, $x(t)$, of the Liouville flow. 

Translation chords and reflection centres are conjugate classical variables 
that define transverse foliations of double phase space. They form the bases for 
complementary representations of quantum operators. The extension of Heisenberg's
uncertainty principle to double phase space, discussed in sections 5 and 6, prevents us from
defining centres and chords simultaneously within the quantum theory. This fact lies
at the root of the dificulty of discussing the near-classical propagation of Wigner
functions, because this region is {\it defined} as the region of small chords,
whereas the Wigner function is constructed in the conjugate centre basis.
In references \cite{RiosOA, OsKon} this problem was partly circumvented by analyzing directly
the evolution of given operators, semiclassically linked to particular Lagrangian surfaces, 
which locally tie a single chord to each centre. However, the classical region of small chords 
for all such surfaces are caustics in both the chord and the Weyl representation, so that improved
uniform approximations become necessary. Here, we have adopted the alternative 
of developing mixed propagators which bypass the uncertainty principle 
because the chord and the centre are each specified at a different instant.
These propagators are privledged representations of the super-operators 
that act on the space of linear operators of Hilbert space.

\ack
We thank Pedro Rios and Raul Vallejos for stimulating discusssions.
Partial financial support from 
Millenium Institute of Quantum Information, PROSUL 
and CNPq is gratefully acknowledged.

\appendix
\section{Centre-centre propagation}

The propagator, $2^L R'_{\x}(\x',t)$, 
that takes a Weyl symbol of an operator into a new Weyl symbol
and thus evolves arbitrary Wigner functions corresponds
classically to a Lagrangian surface in double phase space, $\y'_\x(\x',t)$, that has evolved
from the initial vertical plane, $\x=constant$. This is strictly analogous
to the position propagator, $\langle q_+|\opU_t|q_-\rangle$, in single phase space,
which corresponds to the classical evolution of the vertical plane, $q=q_-$. 
In both cases the initial vertical plane is an extreme nongeneric caustic, 
but the semiclassical position propagator is usually well behaved for a finite 
$t>0$. It might not then be immediately obvious why caustics are inevitable 
in the double space evolution. In other words, why is the evolving Lagrangian surface
allways tangent to a vertical plane at $\y=0$, as shown in Fig.3(b)?

The essential difference lies in the driving Hamiltonians. The typical form of the Hamiltonian
in single phase space corresponding to the Schroedinger equation is $p^2/2+V(q)$, so that
$\dot q=p$. This tilts the initial plane, $q=q_-$, and so breaks the verticallity. Indeed,
to first order in time, the Lagrangian surface is $p=(q-q_-)/t$. In contrast, the double phase
space motion is driven by the Heisenberg Hamiltonian, $I\!\!H'(\x,\y)$, defined by \eref{Heisham}
for an arbitrary single space Hamiltonian $H'(x)$.
Its expansion to lowest order in $\y$ is given by \eref{smallH}, which leads to
an evolution of the $\x$ coordinate that is independent of $\y$. The third order terms in $\y$
contribute quadratic terms to the classical equations of motion which
break the strict verticality, but the vertical tangent remains at $\y=0$.   

In the example at the end of section 8, the Fourier transform of \eref{refl-p3} leads 
immediately to an Airy function for the centre-centre propagator and generically we can
expect uniform approximations based on Airy functions in the case of a single degree of freedom.
This was previously derived for special quantum maps \cite{Berryetal}. For higher dimensions and for
nongeneric cases, higher diffraction catastrophes come into play \cite{Berry-Up}. In contrast, 
the mixed propagator remains caustic free in all these cases.


\section*{Bibliography}
\begin{thebibliography}{99}

\bibitem{Vor76} Voros A 1976 Ann. Inst. Henri Poincar\'e {\bf 26} 31
\bibitem{RiosOA}Rios P~P~M and Ozorio de Almeida A~M 2002 J. Phys. A {\bf 35} 2609.
\bibitem{OsKon} Osborn T~A and Kondratieva MF 2002 J. Phys. A {\bf 35} 5279
\bibitem{Rios04} Rios P~P~M and Ozorio de Almeida AM 2004 J. Geom. Phys. {\bf 51} 404 
\bibitem{Mar91} Marinov MS 1991 Phys. Lett. A {\bf 153} 5
\bibitem{Dittrich} Dittrich T, Viviescas C and Sandoval L 2005 (private communication)
\bibitem{Grossmann} Grossmann A, Commun. Math. Phys. 1976 {\bf 48} 191
\bibitem{Breuer} Breuer H-P and Petruccione F 2002 \textsl{The Theory of Open Quantum Systems}
(Oxford: Oxford University Press)
\bibitem{Kraus} Kraus K 1983 \textsl{States, Effects and Operations}, Lecture Notes in Physics 
{\bf 190} (Berlin: Springer-Verlag) 
\bibitem{Alm98} Ozorio de Almeida A~M 1998 Phys. Rep. {\bf 295}, 265 
\bibitem{AmHu80} Amiet JP and Huguenin P 1980 Helvetica Physica Acta {\bf 53} 377
\bibitem{livro} Ozorio de Almeida A~M 1988 {\it Hamiltonian Systems: Chaos and Quantization}
(Cambridge: Cambridge University Press)
\bibitem{Maslov} Maslov V~P and Fedoriuk M~V  1981 \textsl{Semiclassical Approximation 
in Quantum Mechanics} (Reidel, Dordrecht)
\bibitem{GrossHug} Grossmann A and Huguenin P 1978 Helvetica Physica Acta {\bf 51}, 252
\bibitem{Wigner} Wigner E~P 1932 Phys. Rev. {\bf 40} 749
\bibitem{AlmVal04} Ozorio de Almeida A~M, Vallejos O and Saraceno M 2004
J. Phys. A {\bf 38} 1473 and quant-ph/ 0410129
\bibitem{Giulini} Giulini D, Joos E, Kiefer C, Kupsch J, Stamatescu I-O and Zeh H~D 1996 
\textsl{Decoherence and the Appearance of a Classical World in Quantum Theory} 
 (Springer, Berlin)
\bibitem{Goldstein} Goldstein H 1980 \textsl{Classical Mechanics}, 2nd edition (Addison-Wesley, Reading, M.A.)
\bibitem{Mar79} Marinov M~S 1979 J. Phys. A {\bf 12} 31
\bibitem{Alm90} Ozorio de Almeida A~M 1990 Proc. R. Soc. Lond. A {\bf 431} 403
\bibitem{Berry89a} Berry M~V 1989 Proc. R. Soc. Lond. A {\bf 423} 219 
\bibitem{Marinov} Marinov M~S 1991 Phys. Lett. A {\bf153}, 5 
\bibitem{BroAlm04} Brodier O and Ozorio de Almeida A~M 2004 Phys. Rev. E {\bf 69} 016204  
\bibitem{Littlejohn95} Littlejohn R~G 1995 in \textsl{Quantum Chaos: Between Order and Disorder}, 
edited by Casati G and Chirikov B, 343-404 (Cambridge University Press, Cambridge)
\bibitem{Chountasis} Chountasis S and Vourdas A 1998 Phys Rev. A {\bf 58} 1794
\bibitem{Moyal} Moyal J~E 1949 Proc. Camb. Phil. Soc. Math Phys. Sci. {\bf 45} 99
\bibitem{Dodonov86} Dodonov V~V and Man'ko V~I Physica 1986 {\bf 137}A 306
\bibitem{Alm84} Ozorio de Almeida A~M 1984 Rev. Bras. Fis. {\bf 14} 62 
\bibitem{Ber77} Berry M~V 1977 Phil. Trans. Roy. Soc A {\bf 287} 237-71
\bibitem{AlmHan82} Ozorio de Almeida A~M and Hannay J 1982 Ann. Phys. {\bf 138} 115
\bibitem{Tabor} Tabor M 1983 Physica {\bf 6D} 195
\bibitem{Gutzwiller} Gutzwiller M~C 1990 \textsl{Chaos in Classical and Quantum Mechanics}
 (Springer, New York)
\bibitem{Berry-Up} Berry M~V and Upstill C 1980 Prog. Opt. {\bf 18} 149
\bibitem{Berryetal} Berry M~V, Balazs N~L, Tabor M and Voros A 1979 Ann. Phys. NY {\bf 122} 26

\end{thebibliography}
\end{document}